\documentclass[DIV12]{scrartcl}

\usepackage{hyperref}
\usepackage[utf8]{inputenc}
\usepackage{tabularx}
\usepackage{setspace}
\usepackage{graphicx}
\usepackage{booktabs}
\usepackage{amsmath}

\hypersetup{
	colorlinks=true,
	linkcolor=black,
	citecolor=black,
	urlcolor=black
}

\date{}
\usepackage{authblk}
\usepackage{blindtext}

\usepackage{natbib}
\title{Transforming Musical Signals through a Genre Classifying Convolutional Neural Network}

\usepackage{fancyhdr}
\begin{document}

\author[1]{Shijia Geng\thanks{s.geng@umiami.edu}}
\author[1]{Gang Ren\thanks{gang.ren.music@gmail.com}}
\author[1, 2]{Mitsunori Ogihara\thanks{ogihara@cs.miami.edu}}

\affil[1]{\small Center for Computational Science, University of Miami, Coral Gables, FL, USA}
\affil[2]{\small Department of Computer Science, University of Miami, Coral Gables, FL, USA}

\maketitle
\thispagestyle{fancy} 

\begin{abstract}
Convolutional neural networks (CNNs) have been successfully applied on both discriminative and generative modeling for music-related tasks. For a particular task, the trained CNN contains information representing the decision making or the abstracting process. One can hope to manipulate existing music based on this ``informed" network and create music with new features corresponding to the knowledge obtained by the network. In this paper, we propose a method to utilize the stored information from a CNN trained on musical genre classification task. The network was composed of three convolutional layers, and was trained to classify five-second song clips into five different genres. After training, randomly selected clips were modified by maximizing the sum of outputs from the network layers. In addition to the potential of such CNNs to produce interesting audio transformation, more information about the network and the original music could be obtained from the analysis of the generated features since these features indicate how the network ``understands" the music.

\bigskip


\end{abstract}

\section{Introduction}
Recent studies show that CNNs provide promising performance when applied to the music-related fields such as genre classification, instrument recognition, and music composing ~\citep{dieleman2011audio}.  A CNN trained on a particular music-related task ``learns" one type of knowledge about music and it has the potential to be applied to reshaping music or audio recording. It can be viewed as an audio counterpart of the deepdream project which adjusts images to enhance patterns based on CNN trained on visual recognition tasks ~\citep{mordvintsev2015inceptionism}.

\section{Methods}
In this study, a CNN with three convolutional layers was trained to classify music into five genres (alternative, electronica, pop, rap, and rock). The private datasets for training and testing were prepared from 23,639 song files. The signals obtained from the audio files were downsampled to 8,000Hz and separated into five-second (40,000 data point) clips. Eventually, a total of 139,500 audio clips were prepared.

The major components of the network include an input layer, three convolutional layers, and an output layer. The shapes of the input and output layers are $1 \times 40,000$ and $1 \times 5$, respectively. The convolutional layers hold 16 filters each, having length 8, 32, and 128, respectively. The stride length for each filter is 8. The activation function for each node in the convolutional layers is the rectify function. In addition, batch normalization was applied on the intermediate result generated by each convolutional layer. 

The optimization approach used in the classification task was mini-batch gradient descent, which took a subset of data in each iteration and updated the parameters based on the gradient of the aggregated loss. The specific update method applied was based on \textit{Nesterov's accelerated gradient descent}. During training, twenty epochs were applied and the training loss and training time for each epoch were recorded. 

The computational goal for the modification procedure was to modify the input clip in order to maximize the objective function which is the sum of a particular convolutional layer or all convolutional layers. The optimization method chosen to maximize the objective function was gradient ascent. Each modified clip was normalized and saved as a 16-bit audio file in the linear pulse-code modulation waveform audio file format. 

\section{Results and Discussion}
The classification accuracies for the five genres, alternative, electronica, pop, rap and rock music, are respectively 27.6\%, 45.3\%, 34.2\%, 84.3\%, and 51.2\%. The results indicate that the proposed CNN has better performances for the genres with outstanding characteristic such as rap and rock. However, its ability to distinguish genres with uncertain features such as alternative and pop is weak. A more complicated network architecture may be required in order to overcome the weakness.

The resulting audio files are available at ~\citep{ModifiedAudio}. The ones based on lower layers tend to develop some attack-like features, while the ones based on higher layers have clear harmonics pattern. More constraints might be necessary to reduce the noises in the modification.

%\blindtext[1]

\bibliography{dl_music_refs}

\end{document}